\documentclass[11pt]{revtex4-1}

\usepackage{graphicx}

\begin{document}

\title{A Study of Universal Thermodynamics in Massive Gravity: Modified Entropy on the Horizons}

\author{Subhajit Saha\footnote {subhajit1729@gmail.com}}
\author{Saugata Mitra\footnote {saugatamitra20@gmail.com}}
\author{Subenoy Chakraborty\footnote {schakraborty.math@gmail.com}}

\affiliation{Department of Mathematics, Jadavpur University, Kolkata 700032, West Bengal, India.}

\begin{abstract}

Universal thermodynamics for FRW model of the Universe bounded by apparent/event horizon has been considered for massive gravity theory. Assuming Hawking temperature and using the unified first law of thermodynamics on the horizon, modified entropy on the horizon has been determined. For simple perfect fluid with constant equation of state, generalized second law of thermodynamics and thermodynamical equilibrium have been examined on both the horizons. \\\\
Keywords: Massive gravity theory, Modified entropy, Generalized second law of thermodynamics, Thermodynamical equilibrium \\\\
PACS Numbers: 04.50.Kd, 98.80.-k, 95.30.Sf

\end{abstract}

\maketitle

\section{Introduction}

Recent astronomical data from Type Ia Supernovae (SNe Ia) \cite{Riess1}, Cosmic Microwave Radiation Background (CMB) \cite{Spergel1} and Large Scale Structure (LSS) \cite{Eisenstein1} provides ample evidences that the present Universe is undergoing an expansion which is rather accelerated. The cause for this acceleration is still debated. A group of cosmologists have been trying to incorporate this late time acceleration into standard cosmology by the introduction of an exotic matter which has been dubbed "Dark energy" (DE). This hypothetical matter is thought to have a huge negative pressure, thus causing the Universe to accelerate. However, in spite of extensive research, the nature of DE is still a mystery. 

Another group of cosmologists are of the opinion of a modified theory of gravity$-$a modification of Einstein's general theory of relativity. A common and widely used modified gravity theory is $f(R)-$gravity, where the Lagrangian density $R$ (Ricci scalar) in the Einstein Hilbert action is replaced by an arbitrary function of $R$, i.e., $f(R)$ (see \cite{Sotiriou1} and the references therein). One can find many other modified gravity theories (higher dimensional theories as well) in the literature. These modified theories \cite{Capozziello1, Nojiri1, Starobinsky1, Nojiri2, Nojiri3, Bengochea1} are considered as gravitational alternatives for DE and might serve as dark matter \cite{Sobouti1}. In the present work, we consider massive gravity theory as a modified theory of gravity and examine the thermodynamical behaviour both at the apparent and event horizons. The paper is organized as follows. Section II describes the basic features of universal thermodynamics. The basic equations in massive gravity theory are presented in Section III. Thermodynamics in massive gravity has been discussed in Section IV. Finally, a brief discussion and final comments are given in Section V.

\section{Universal Thermodynamics: Basic Features}

A lot of research in recent years has been carried out in Universal thermodynamics, mostly with an apparent horizon as the boundary. In 2006, Wang {\it et al.} \cite{Wang1} made a comparative study of the two horizons (apparent and event) by examining the validity of the thermodynamical laws for DE fluids and concluded that the Universe bounded by an apparent horizon is a Bekenstein system whereas a cosmological event horizon is unphysical from the thermodynamical point of view. However, it has been shown \cite{Mazumder1} that the generalized second law of thermodynamics holds (in any gravity theory) with some reasonable restrictions for Universe bounded by an event horizon under the assumption that the first law holds. Further, a modified form \cite{Chakraborty1} of the Hawking temperature has been identified recently using which, it has been possible to show \cite{Chakraborty2} the validity of both the first and the second laws of thermodynamics for Universe bounded by an event horizon and the results obtained are independent of the fluid taken.  

According to thermodynamical concepts, the entropy of an isolated macroscopic physical systems never decrease because such systems always evolve toward thermodynamic equilibrium, a state with maximum entropy (consistent with the constraints imposed on the system). Thus, for a universe filled with a fluid and bounded by a horizon, the (generalized) second law of thermodynamics and the thermodynamical equilibrium respectively take the forms \cite{Pavon1, Saha1}
\begin{equation} \label{tecases}
\dot{S}_h+\dot{S}_{fh} \geq 0~~and~~\ddot{S}_h+\ddot{S}_{fh} < 0, 
\end{equation}
where $S_{h}$ and $S_{fh}$ are the entropies of the horizon and the fluid within it, respectively. In order to determine $\dot{S}_h$, we shall use the Clausius relation 
\begin{equation}
T_h dS_h=\delta Q_h=-dE_h
\end{equation}
and $\dot{S}_{fh}$ can be obtained from the Gibb's relation \cite{Mazumder1, Izquierdo1} which is given by
\begin{equation} \label{fhe}
T_{f} dS_{fh}=dE_f+pdV_h,
\end{equation}
where $E_h$ is the energy flow across the horizon, $E_f=\rho V_h$ is the total energy of the fluid bounded the horizon, $V_h=\frac{4}{3}\pi R_{h}^{3}$ is the volume of the fluid and ($T_h$, $T_f$) are the temperatures of the horizon and fluid inside it, respectively.

On the other hand, in the context of universal thermodynamics, the thermodynamical aspects of dynamical black hole (BH) \cite{Hayward1, Hayward2} was studied by Hayward. THe concept of trapping horizon was introduced in 4D Einstein gravity for non-stationary spherically symmetric spacetimes. As a result, the Einstein field equations are equivalent to the unified first law (UFL). However, the first law of thermodynamics can be derived by projecting the UFL along any tangential direction ($\xi$) to the trapping horizon \cite{Cai1, Akbar1, Cai2} and the Clausius relation of the dynamical BH is written as
\begin{equation}
\langle A\psi , \xi \rangle =\frac{\kappa}{8\pi G}\langle dA, \xi \rangle ,
\end{equation}
where $A$ is the area of the horizon and the energy flux $\psi$ is termed as energy supply vector.

Further, in view of universal thermodynamics, our Universe is assumed to be a non-stationary gravitational system while from the cosmological aspect, the homogeneous and isotropic FRW Universe may be considered as dynamical spherically symmetric spacetime. As a result, there is only inner trapping horizon which coincides with the apparent horizon. So it would be interesting to have thermodynamical analysis using UFL. A first step along this line was taken in 2009 by Cai and Kim \cite{Cai3}. They were able to derive the Friedmann equations with arbitrary spatial curvature starting with the fundamental relation $\delta Q=TdS$ at the apparent horizon of the FRW Universe. They have considered Hawking temperature ($T_H$) and Bekenstein entropy ($S_B$) on the apparent horizon as
\begin{equation}
T_H=\frac{1}{2\pi R_A}~,~~~S_B=\frac{\pi R_{A}^{2}}{G},
\end{equation}
with $R_A$ as the radius of the apparent horizon. Moreover, they were able to show the equivalence between the thermodynamical laws and modified Einstein equations in Gauss-Bonnet gravity and more general Lovelock gravity. Then Cai and others \cite{Cai1, Akbar1, Cai2} examined the UFL in the background of modified gravity theories, namely Lovelock gravity, scalar-tensor theory \cite{Cai1} and brane-world scenario \cite{Cai2}. However, in $f(R)$ gravity theory, one needs entropy production term \cite{Eling1} for the fulfilment of Clausius relation. Subsequently, thermodynamical laws have been studied \cite{Karami1, Wu1} in $f(R)$ (generalized $f(R)$) gravity with a modified version of the horizon entropy. Very recently, we have modified horizon entropy \cite{Mitra1} in such a manner that Clausius relation is automatically satisfied. The present work is an extension of that work in massive gravity theory.

The homogeneous and isotropic FRW model is described by its line element as 
\begin{eqnarray}
ds^2 &=& -dt^2 + \frac{a^2(t)}{1-kr^2}dr^2 + R^2 d\Omega_{2}^2
\nonumber
\\
&=& h_{ab}dx^a dx^b + R^2 d\Omega_{2}^2,
\end{eqnarray}
where $R=ar$ is the area radius, $h_{ab}=diag(-1,\frac{a^2}{1-kr^2})$ is the metric on 2-space $(x^0=t,x^1=r)$ and $k=0,\pm1$ denotes the curvature scalar. Also in double null coordinates ($\xi ^{\pm}$), the above FRW line element takes the form \cite{Cai1}
\begin{equation}
ds^2=-2d\xi^+ d\xi^- +R^2 d\Omega_{2}^2,
\end{equation}
where 
\begin{equation}
\partial _{\pm}=\frac{\partial}{\partial \xi ^{\pm}}=-\sqrt{2}\left(\frac{\partial}{\partial t} \mp \frac{\sqrt{1-\kappa r^2}}{a}\frac{\partial}{\partial r}\right)
\end{equation}
are future pointing null vectors. \\\\
The trapping horizon (denoted by $R_T$) is defined as $\partial_+ R|_{R=R_T} =0$, i.e.,
\begin{equation}
R_T=\frac{1}{\sqrt{H^2+\frac{k}{a^2}}}=R_A
\end{equation}
is the radius of the trapping horizon.\\\\
The surface gravity is defined as 
\begin{equation}
\kappa=\frac{1}{2\sqrt{-h}}\partial_a(\sqrt{-h}h^{ab}\partial_b R),
\end{equation}
so for any horizon having radius $R_h$, we have
\begin{equation}
\kappa _h=-\left(\frac{R_h}{R_A}\right)^2\left(\frac{1-\frac{\dot{R}_A}{2HR_A}}{R_h}\right),
\end{equation}
and
\begin{equation} \label{sgah}
\kappa _A=-\frac{1-\epsilon}{R_A},
\end{equation}
with $\epsilon =\frac{\dot{R}_A}{2HR_A}$ is the surface gravity at the apparent horizon.

In most of the modified gravity theories, the Einstein field equations in FRW model can be written in the form of modified Friedmann equations as
\begin{equation} \label{mfe1}
H^2+\frac{k}{a^2}=\frac{8\pi G}{3}\rho_t
\end{equation}
and
\begin{equation} \label{mfe2}
\dot{H}-\frac{k}{a^2}=-4\pi G(\rho_t+p_t),
\end{equation}
where $\rho_t=\rho +\rho_e$ and $p_t=p+p_e$ are the total energy density and the thermodynamic pressure, ($\rho$, $p$) are the corresponding quantities for the matter distribution while ($\rho _e$, $p_e$) are termed as effective quantities due to the curvature (or other) contributions.

According to Cai \cite{Cai1}, the energy supply vector $\psi$ and the work density $W$ are defined as \cite{Hayward1, Hayward2, Cai1, Akbar1, Cai2}
\begin{equation}
\psi_a=T_a^b\partial_b R+W\partial_a R~,~~~W=-\frac{1}{2}T^{ab}h_{ab}.
\end{equation}
Now, for the above modified Friedmann equations (Eqs. (\ref{mfe1}) and (\ref{mfe2})), the explicit form of $W$ and $\psi$ are
\begin{eqnarray}
W &=& \frac{1}{2}(\rho_t-p_t)=\frac{1}{2}(\rho -p) + \frac{1}{2}(\rho_e-p_e) \nonumber \\
&=& W_m+W_e
\end{eqnarray}
and 
\begin{eqnarray} \label{psi}
\psi &=& \psi _m + \psi _e \nonumber \\
&=& \left\lbrace -\frac{1}{2}(\rho +p)HRdt+\frac{1}{2}(\rho +p)adr \right\rbrace + \left\lbrace -\frac{1}{2}(\rho_e +p_e)HRdt+\frac{1}{2}(\rho_e +p_e)adr \right\rbrace .
\end{eqnarray}
It should be noted that only the pure matter energy supply $A\psi _m$ will give the heat flow $\delta Q$ in the Clausius relation when projected on the horizon. Also the ($0$,$0$) component of the modified Einstein equations (i.e., Eq. (\ref{mfe1})) can be written as the unified first law \cite{Hayward1}
\begin{equation} \label{ufl}
dE =A\psi +WdV
\end{equation}
with $V=\frac{4}{3}\pi R^3$ as the volume of the sphere of radius $R$. Further, any vector $\xi$ tangential to the apparent horizon can be expressed in terms of the null vectors $\partial _{\pm}$ as
\begin{equation}
\xi =\xi _+ \partial _+ + \xi _- \partial _- .
\end{equation} 
As the trapping horizon is characterized by $\partial _+R_T=0$, so on the marginal sphere, $\xi (\partial _+R_T)=0$, which gives
\begin{equation}
\frac{\xi _+}{\xi _-}=-\frac{\partial _- \partial _+ R_T}{\partial _+ \partial _+ R_T}.
\end{equation}
As for the present FRW model $R_T=R_A$, so we have
\begin{equation}
\partial _- \partial _+ R_A=\frac{4}{R_A}(1-\epsilon)~,~~~\partial _+ \partial _+ R_A=-\frac{4\epsilon}{R_A}.
\end{equation}
Hence in ($r$,$t$) coordinates, the tangent vector $\xi$ can be written as \cite{Cai1}
\begin{equation}
\xi =\frac{\partial}{\partial t}-(1-2\epsilon)H_r \frac{\partial}{\partial r}.
\end{equation}
Thus projecting the above UFL (i.e., Eq. (\ref{ufl})) along $\xi$, the true first law of thermodynamics at the apparent horizon takes the form \cite{Cai1, Akbar1, Cai2}
\begin{equation} \label{uflah}
\langle dE, \xi \rangle = \frac{\kappa}{8\pi G}\langle dA, \xi \rangle +\langle WdV, \xi \rangle .
\end{equation}
As heat flow $\delta Q$ corresponds to pure matter energy supply $A \psi _m$, projecting on the apparent horizon so form Eq. (\ref{uflah}), we obtain \cite{Mitra1}
\begin{equation}
\delta Q=\langle A\psi _m, \xi \rangle = \frac{\kappa _A}{8\pi G}\langle dA, \xi \rangle -\langle A\psi _e, \xi \rangle .
\end{equation}
Now using Eqs. (\ref{psi}) and (\ref{sgah}), the explicit form of $\delta Q$ is 
\begin{equation} \label{Q}
\delta Q=-\frac{2\epsilon (1-\epsilon)}{G}HR_A +A(1-\epsilon)HR_A (\rho _e+p_e)
\end{equation}
with Hawking temperature on the apparent horizon as
\begin{equation}
T_A=\frac{\kappa _A}{2\pi}=\frac{(1-\epsilon)}{2\pi R_A}.
\end{equation}
Hence Eq. (\ref{Q}) can be rewritten as \cite{Mitra1}
\begin{equation}
\delta Q=\langle A\psi _m, \xi \rangle = T_A \langle \frac{8\pi R_A}{4G}dR_A-8 \pi ^2 H R_{A}^{4}(\rho _e+p_e)dt, \xi \rangle  .
\end{equation}
On comparison with Clausius relation $\delta Q=TdS$, the differential form of the entropy on the apparent horizon takes the form \cite{Mitra1}
\begin{eqnarray}
dS_A &=& \frac{2\pi R_A}{G}dR_A-8\pi ^2 HR_{A}^{4}(\rho _e+p_e)dt \\
i.e.,~~~~S_A &=& \frac{A_A}{4G}-8\pi ^2 \int HR_{A}^{4}(\rho _e+p_e)dt. \label{sa}
\end{eqnarray}
This shows that the entropy on the apparent horizon is the usual Bekenstein entropy with a correction term (in integral form).

For the event horizon, as $d\xi ^{\pm}=dt \mp adr$, one form along the normal direction, so the tangent vector to the event horizon can be taken as
\begin{equation}
\partial _{\pm}=-\sqrt{2}(\partial _t \mp \frac{1}{a}dr),
\end{equation}
i.e., one can choose
\begin{center}
$\xi =\frac{\partial}{\partial _t}-\frac{1}{a}\frac{\partial}{\partial _r}$
\end{center}
as the tangential vector to the surface of the event horizon. Thus, proceeding as above, the entropy on the event horizon can be written as \cite{Mitra1}
\begin{equation} \label{se}
S_E=\frac{A_E}{4G}-4\pi ^2 \int \left(\frac{R_{A}^{2}R_E}{1-\epsilon}\right)\left(\frac{HR_E+1}{HR_E-1}\right) (\rho _e+p_e)dR_E
\end{equation}
which again shows that the leading term for entropy is the usual Bekenstein entropy.

\section{Basic Equations in Massive Gravity}

In recent times, a new modified theory of gravity has been constructed by adding a small mass to the graviton. In classical field theory, there arises a basic question of whether a consistent extension of general relativity (GR) by a mass term is possible or not. Fierz and Pauli (FP) initiated this attempt to construct a theory of gravity with massive graviton \cite{Fierz1}. They added a quadratic mass term $m^2$ ($h_{\mu \nu} h^{\mu \nu}-h^2$) to the action (for linear gravitational perturbations) and as a result there was a violation of gauge invariance in GR. Further, the linear theory with FP-mass does not match with GR in the zero mass limit and this leads to the contradiction with solar system tests due to the vDVZ discontinuity \cite{Van Dam1, Zakharov1}. However, introduction of nonlinear interactions using Vanishtein mechanism \cite{Vainshtein1} might overcome this problem. This idea was subsequently used by Dvali, Gabadadze and Porrati \cite{Dvali1} to construct a higher dimensional model of massive gravity which admits a self-accelerating solution with dust matter and the theory modifies GR at the cosmological scale.

On the other way, the Bouldware-Deser (BD) ghost \cite{Boulware1} in the formulation with St$\ddot{u}$ckelberg field can be eliminated by adding nonlinear interactions with higher derivatives, order by order in perturbation theory. Recently, de Rham, Gabadadze and Tolley (dGRT) formulated a nonlinear massive gravity theory which is free from BD ghost \cite{de Rham1}. Subsequently, various cosmological solutions \cite{D'Amico1, Gong1, Gumrukeuoglu1, Gratia1, Kobayashi1, Langlois1, Volkov1, Von Strauss1} have been evaluated using dGRT-massive gravity theory. To ensure that no ghost appears in the decoupling limit, the total action takes the form \cite{de Rham1}
\begin{equation}
S=\frac{1}{16\pi G} \int {d^4 x \sqrt{-g} [R+m_{g}^{2} U]+S_m},
\end{equation}
where $m_g$ stands for graviton mass and the nonlinear higher derivative term $U$ corresponding to massive graviton has the expression [19,27]
\begin{equation}
U=U_2+\alpha _3 U_3 +\alpha _4 U_4
\end{equation}
with
\begin{equation}
U_2=[\kappa]^2 - [\kappa ^2],
\end{equation}
\begin{equation}
U_3=[\kappa]^3-3[\kappa][\kappa ^2]+2[\kappa ^3],
\end{equation}
\begin{equation}
U_4=[\kappa]^4-6[\kappa]^2 [\kappa ^2]+8[\kappa ^3][\kappa]-6[\kappa ^4].
\end{equation}
The tensor $\kappa _{\nu}^{\mu}$ is defined as
\begin{equation}
\kappa _{\nu}^{\mu}=\delta _{\nu}^{\mu}-\sqrt{\partial ^{\mu}\phi ^{\alpha} \partial _{\nu} \phi ^{b} f_{ab}},
\end{equation}
where $\phi ^{\alpha}$ stands for Stu\"{c}kelberg field and the reference metric $f_{ab}$ is usually taken to be Minkowskian metric, i.e., $f_{ab}=diag(-1,1,1,1)$. The notation with square bracket represents trace as follows:
\begin{equation}
[\kappa]=tr~\kappa _{\nu}^{\mu}~,~~[\kappa]^2=(tr~\kappa _{\nu}^{\mu})^2~,~~[\kappa ^2]=tr~\kappa _{\nu}^{\mu} \kappa _{\lambda}^{\nu}
\end{equation}
and so on.

For convenience, if the unitary gauge $\phi ^a(x)=x^\mu \delta _{a}^{\mu}$ is chosen, then the metric tensor stands for the observable describing the 5 degrees of freedom of the massive graviton. However, for the Minkowskian reference metric, the theory does not even admits any nontrivial flat homogeneous and isotropic (FLRW) solution \cite{De Felice1}, so in the present work, we choose the reference metric $f_{ab}$ as the de Sitter metric. As a result, it is possible to have the flat, open or closed cosmologies by suitable slicing of de Sitter model \cite{Gong2} and also this choice of reference metric eliminates the problem of 'no-go' theorem \cite{D'Amico1}. Hence the reference metric is taken as
\begin{equation}
f_{ab}d\phi ^a d\phi ^b =-dT^2+b_{k}^{2}(T)\gamma _{ij}(X)dX^i dX^j
\end{equation}
with
\begin{equation} \label{coeff}
b_{0}(T)=e^{H_c T}~,~~~b_{-1}(T)=H_{c}^{-1}sinh(H_{c}T)~,~~~b_{t}(T)=H_{c}^{-1}cosh(H_{c}T).
\end{equation}
Note that in the limit $H_c \rightarrow 0$, the Minkowski metric is recovered for flat and open cases: $b_{0}=1$, $b_{-1}=T$, while the latter case corresponds to the Milne metric for flat geometry.

Now, choosing the Stu\"{c}kelberg field as
\begin{equation}
\phi ^0=T=f(T)~,~~~\phi ^i=X^i=x^i, 
\end{equation}
one sees that the cosmological symmetries are satisfied and the symmetric ($0$,$2$) tensor
\begin{equation}
\Sigma _{\mu \nu}=f_{ab}\partial _{\mu}\phi ^a \partial _{\nu} \phi ^b
\end{equation}
becomes a homogeneous and isotropic tensor of the form
\begin{equation}
\Sigma _{\mu \nu}=diag(-\dot{f}^2,b_{k}^{2}f(t)\gamma _{ij}).
\end{equation}
Thus the elements of the $\kappa$ matrix has the simple form
\begin{equation}
\kappa _{0}^{0}=1-\zeta _f \frac{\dot{f}}{N}~,~~~\kappa _{i}^{j}=\left(1-\frac{b_{k}(f)}{a}\right)\delta _{i}{j}~,~~~\kappa_{i}^{0}=0~,~~~\kappa _{0}^{i}=0,
\end{equation}
where $\zeta _f$ denotes the sign of '$f$'. Hence the explicit expression for the nonlinear higher derivative term (representing the potential for graviton) in the Lagrangian is
\begin{eqnarray}
L_{mg} &=& \sqrt{-g}U=(a-b_{k}(f))\left[N\left \lbrace a^2(4\alpha _3+\alpha _4 +6)-a(5\alpha _3+2\alpha _4+3)b_{k}(f)+(\alpha _3+\alpha _4)b_{k}^{2}(f)\right\rbrace \right] \nonumber \\
&-& (a-b_{k}(f))\left[\zeta _f \dot{f} \left\lbrace a^2(3\alpha _3+\alpha _4+3)-a(3\alpha _3+2\alpha _4)b_{k}(f)+\alpha _4 b_{k}^{2}(f) +\alpha _4 b_{k}^{2}(f) \right\rbrace \right].
\end{eqnarray}
Considering variation of '$f$' in the Lagrangian, the equation of motion for $f(t)$ can be written as
\begin{equation} \label{ee}
\left[(3\alpha _3+\alpha _4+3)a^2-2(2\alpha _3+\alpha _4+1)ab_{k}(f)+(\alpha _3+\alpha _4)b_{k}^{2}(f)\right]\left(\frac{\dot{a}}{N}-\zeta _f b_{k}^{'}(f)\right)=0.
\end{equation}
Hence, solving for $b_{k}(f)$, one gets
\begin{equation} \label{ft}
b_{k}(f(t))=\mu _{\pm} a(t),
\end{equation} 
where
\begin{equation}
\mu _{\pm}=\frac{1+2\alpha _3+\alpha _4 \pm \sqrt{1+\alpha _3+\alpha _{3}^{2}-\alpha _4}}{\alpha _3+\alpha _4}.
\end{equation}
Thus it is possible to have $f(t)$ from Eq. (\ref{ft}), provided $b_k$ is invertible. Note that in the Minkowskian domain (i.e., $f_{ab}=\eta _{ab}$), we have $b_{0}(f)=1$ for flat case and hence no solution for '$f$' is possible while there are two branches of solutions for open case ($b_{-1}(f)=f$) \cite{D'Amico1, Gumrukeuoglu1}. Further, from the remaining part of the evolution equation (\ref{ee}), one gets \cite{Fasiello1, Langlois1}
\begin{equation}
\zeta _f b_{k}^{'}(f)=\frac{\dot{a}}{N}.
\end{equation}
So, similarly as before, non-trivial solutions are possible only if inversion of $b_{k}^{'}(f)$ is possible. Note that this case has no analogue with Minkowskian reference metric because there does not exist any solution for both flat and open cases. However, choosing $b_{k}$'s from Eq. (\ref{coeff}), one gets an explicit solution for $f(t)$ as
\begin{equation} \label{ft1}
f(t)=H_{c}^{-1}ln \left(\frac{H(t)a(t)}{H_c}\right),
\end{equation}
where $H=\frac{1}{N}\frac{\dot{a}}{a}$ is the usual Hubble parameter and $N$ is the lapse function (which sets to unity in the subsequent steps).

Thus for the present FLRW metric, the usual Einstein-Hilbert (EH) component has the explicit form
\begin{equation}
L_{EH}=-\frac{3\dot{a}^2 a}{N}+3\kappa Na
\end{equation}
and in addition to the massive gravity part, there is Lagrangian $L_m$ corresponding to ordinary cosmological matter. So the variation of the total Lagrangian with respect to the lapse function $N$ and the scale factor $a$ gives the first and second Friedmann equations as
\begin{equation}
3\left(H^2+\frac{\kappa}{a^2}\right)=8\pi G (\rho +\rho _e)
\end{equation}
and
\begin{equation}
\dot{H}-\frac{\kappa}{a^2}=-4\pi G(\rho +p_m+\rho _e+p_e),
\end{equation}
where (as in the previous section) ($\rho$, $p$) are the energy density and thermodynamic pressure of the cosmic fluid and ($\rho _e$, $p_e$) are the effective energy density and thermodynamic pressure due to contribution from the massive gravity part with explicit expressions
\begin{eqnarray}
\rho _e &=& \frac{m_{g}^{2}}{8\pi G a^3}(b_{k}(f)-a)\left\lbrace (4\alpha _3+\alpha _4+6)a^2-a(5\alpha _3+2\alpha _4+3)b_{k}(f)+(\alpha _3+\alpha _4)b_{k}^{2}(f)\right\rbrace \nonumber \\
p_e &=& \frac{m_{g}^{2}}{8\pi G a^2}\left[\left\lbrace 4\alpha _3+\alpha _4+6-(3\alpha _3+\alpha _4+3)\dot{f}\right\rbrace a^2 -2\left\lbrace 3\alpha _3+\alpha _4+3-(2\alpha _3+\alpha _4+1)\dot{f}\right\rbrace ab_{k}(f)\right] \nonumber \\
&+& \frac{m_{g}^{2}}{8\pi G a^2}\left[\left\lbrace 1+2\alpha _3+\alpha _4-(\alpha _3+\alpha _4)\dot{f}\right\rbrace b_{k}^{2}(f)\right].
\end{eqnarray}
Now, based on evidences from the WMAP data \cite{Bennet1}, in the present work, we shall restrict ourselves to flat FLRW model of the Universe. So using solution (\ref{ft1}) for $f(t)$, the simplified form of $\rho _e$ and $p_e$ are the following:
\begin{eqnarray}
\rho _e &=& \frac{m_{g}^{2}}{8\pi G} \left[-6\alpha +9\beta \frac{H}{H_c}-3\gamma \frac{H^2}{H_{c}^{2}}+3\delta \frac{H^3}{H_{c}^{3}}\right] \nonumber \\ 
p_e &=& -\rho _e+\frac{m_{g}^{2}}{8\pi G} \frac{\dot{H}}{H^2} \frac{H}{H_c}\left[-3\beta +2\gamma \frac{H}{H_c}-3\delta \frac{H^2}{H_{c}^{2}} \right], \label{pnew}
\end{eqnarray}
where
\begin{equation}
\alpha = 1+2\alpha _3+2\alpha _4~,~~~\beta = 1+3\alpha _3+4\alpha _4~,~~~\gamma = 1+6\alpha _3+12\alpha _4~,~~~\delta = \alpha _3+4\alpha _4.
\end{equation}
Thus the present massive gravity theory contains three free parameters namely $m_g$, $\alpha _3$ and $\alpha _4$ and it should be noted that there is no longer any contribution of massive graviton to the energy density (i.e., $\rho _e=0$) if $H=H_c$.

\section{A Thermodynamical Study in Massive Gravity Theory}

This section deals with universal thermodynamics of massive gravity theory in the background of flat FLRW model. So the modified Friedmann equations take the form
\begin{eqnarray}
H^2 &=& \frac{8\pi G}{3}(\rho +\rho _e) \nonumber \\
\dot{H} &=& -4\pi G (\rho +p +\rho _e+p_e).
\end{eqnarray}

Now the expression for entropy on the apparent and the event horizon (after some algebra) can be obtained in massive gravity theory (for details of calculation, one may refer to Appendix A) from Eqs. (\ref{sa}) and (\ref{se}) as
\begin{equation} \label{samg}
S_A=\frac{A_A}{4G}-\frac{\pi m_{g}^{2}}{GH_c}\left[\beta R_{A}^{3}-\frac{\gamma}{H_c}R_{A}^{2}+\frac{3\delta}{H_c^{2}}R_A\right]
\end{equation}
and
\begin{equation} \label{semg}
S_E=\frac{A_E}{4G}-\frac{\pi m_{g}^{2}}{2GH_c}\int \left[\frac{2R_E(HR_E+1)}{2-\dot{R}_A}\left\lbrace -\frac{3\beta}{H^3}+\frac{2\gamma}{H_c}\frac{1}{H^2}-\frac{3\delta}{H_c^{2}}\frac{1}{H}\right\rbrace\right]dH
\end{equation}
respectively. One should note that the first term in both the above expressions is the usual Bekenstein entropy for the corresponding horizon, $A_A$ and $A_E$ being the area of the horizons.

\begin{figure}
\begin{minipage}{0.4\textwidth}
\includegraphics[width=1.0\linewidth]{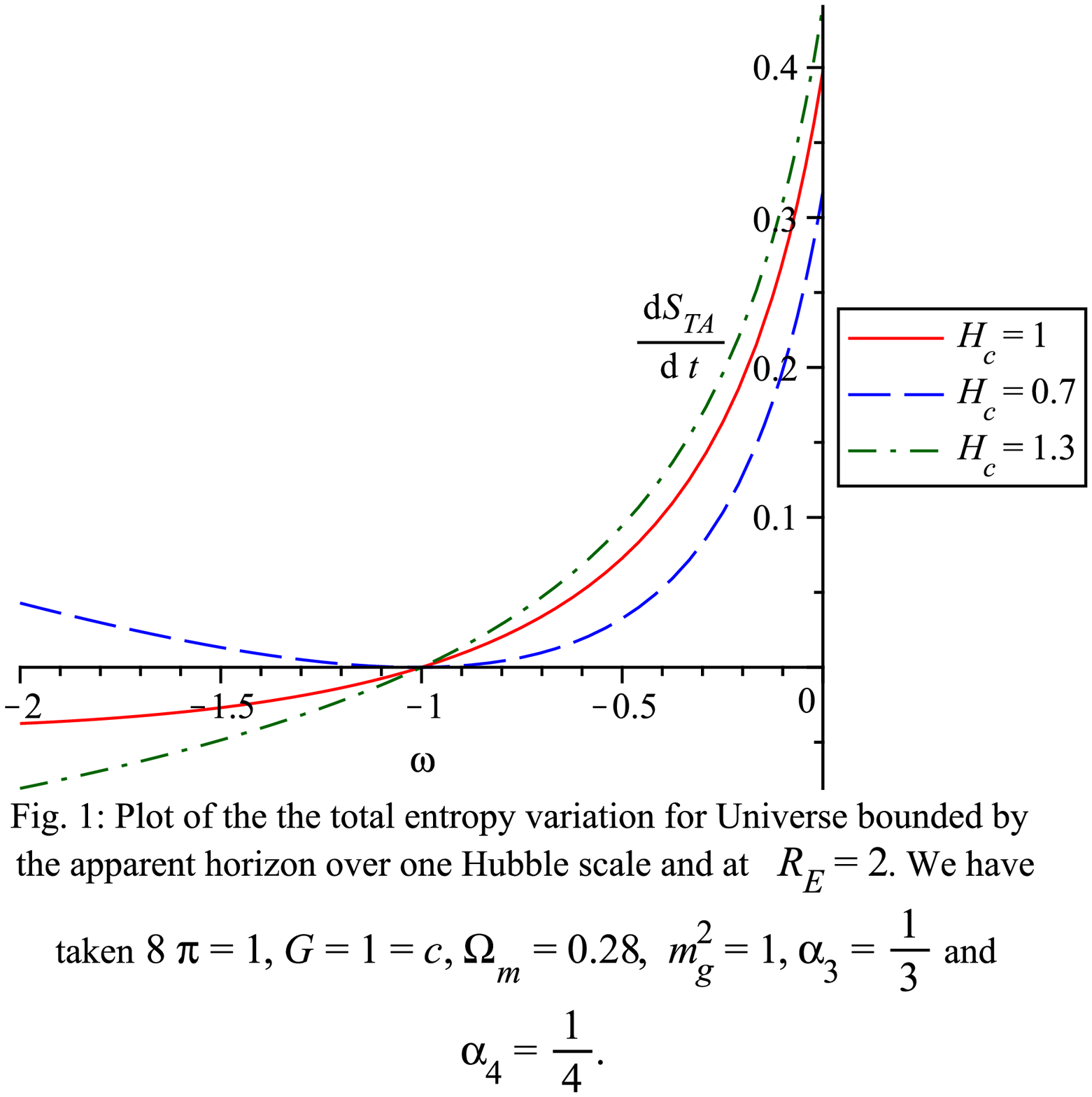}
\end{minipage}
\begin{minipage}{0.4\textwidth}
\includegraphics[width=1.0\linewidth]{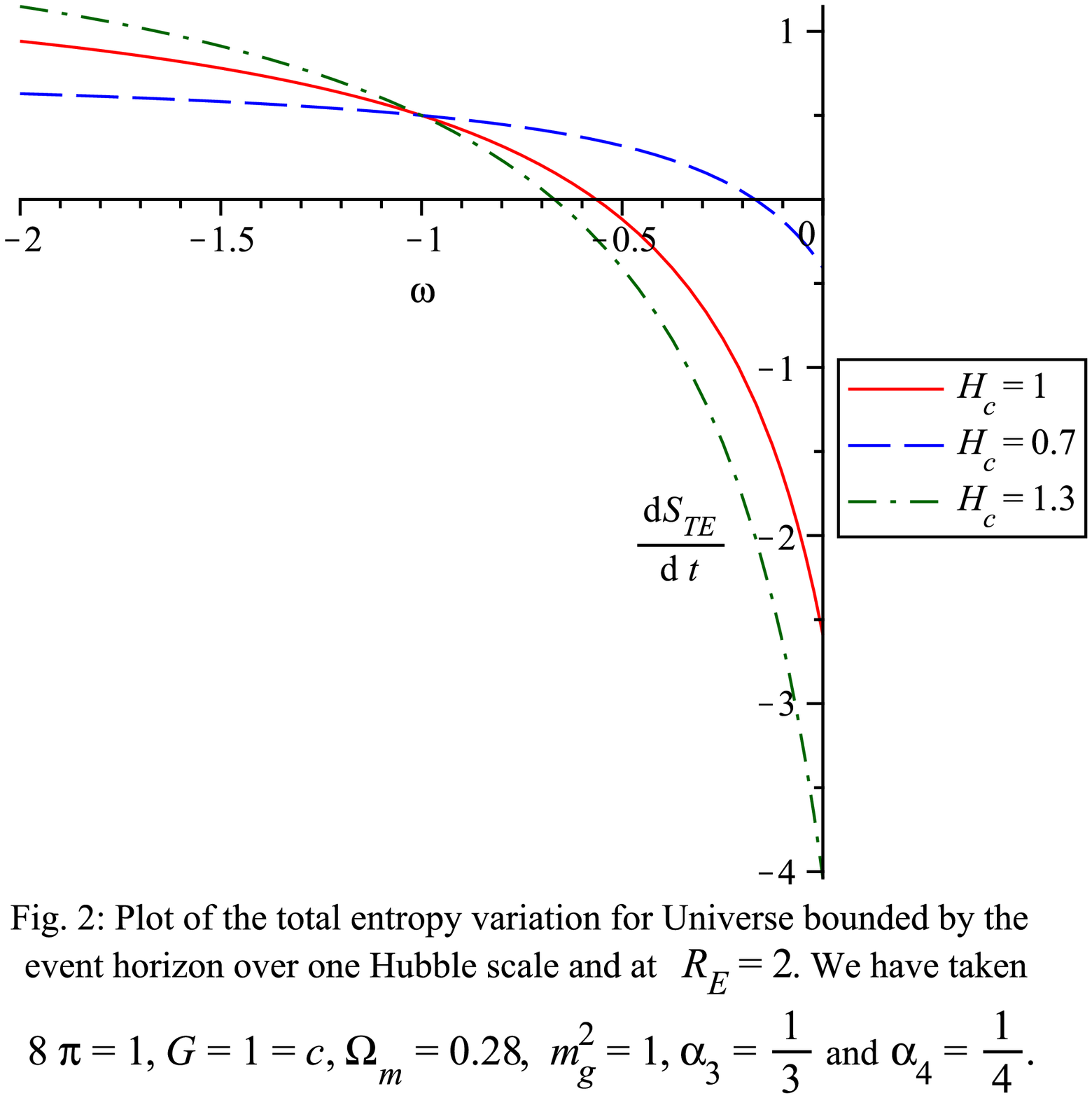}
\end{minipage}
\end{figure}

The expression for the entropy variation of the fluid bounded by any horizon (given in Eq. (\ref{fhe})) can be expressed in a general form as (for details of calculation, one may refer to Appendix B)
\begin{equation} \label{dfh}
\dot{S}_{fh}=\frac{4\pi R_{h}^{2}}{T_f} (\rho +p)(\dot{R}_h-HR_h).
\end{equation}
For a flat model of the Universe, the radii of the two horizons (namely apparent and event) are related by the inequality $R_A=\frac{1}{H}<R_E$. THe thermodynamical parameters namely (Bekenstein) entropy and (Hawking) temperature on the apparent horizon are given by
\begin{equation} \label{sa1}
S_A=\frac{\pi R_{A}^{2}}{G}~~and~~T_A=\frac{1}{2\pi R_A}.
\end{equation}
However, following Ref. \cite{Chakraborty1}, the entropy and the temperature on the event horizon are written as
\begin{equation} \label{se1}
S_E=\frac{\pi R_{E}^{2}}{G}~~and~~T_E=\frac{R_E}{2\pi R_{A}^{2}}=\frac{H^2 R_E}{2\pi}.
\end{equation}

Thus, taking derivatives of Eqs. (\ref{samg}) and (\ref{semg}), adding them respectively to Eq. (\ref{dfh}) and substituting the expressions of $T_A$ and $T_E$ from Eqs. (\ref{sa1}) and (\ref{se1}), one can evaluate the total entropy variation for Universe bounded by the apparent and the event horizons as (considering the temperature of the horizons and that of the fluid inside them to be equivalent)
\begin{equation} \label{dsta}
\dot{S}_{TA}=\frac{3\pi}{2G}(1+\omega)\left[2\left\lbrace 1+16\pi G \Omega _m \left(\frac{1+3\omega}{1-3\omega}\right)\right\rbrace R_A-\frac{m_{g}^{2}}{H_c}\left(3\beta R_{A}^{2}-\frac{2\gamma}{H_c}R_A+\frac{3\delta}{H_{c}^{2}}\right)\right]
\end{equation}
and
\begin{eqnarray} \label{dste}
\dot{S}_{TE}&=&\frac{\pi}{G}\Big[2R_E(HR_E-1)-\frac{3m_{g}^{2}}{H_c}\left(\frac{1+\omega}{1-3\omega}\right)R_E(HR_E+1)\left\lbrace 3\beta R_A-\frac{2\gamma}{H_c}+\frac{3\delta}{H_{c}^{2}}\frac{1}{R_A} \right\rbrace \nonumber \\
&-& 96\pi G \Omega _m\left(\frac{1+\omega}{1-3\omega}\right)R_E\Big]
\end{eqnarray}
respectively. We have assumed that the Universe is filled with a perfect fluid having constant equation of state, i.e., $p=\omega \rho$ where $\omega$ is a constant and $\Omega _m=\frac{\rho}{3H^2}$ is the usual density parameter. The velocities of the apparent and the event horizons are given by
\begin{equation}
v_A=\dot{R}_A=-\frac{\dot{H}}{H^2}=\frac{3}{2}(1+\omega)
\end{equation}
and
\begin{equation}
v_E=\dot{R}_E=(HR_E-1).
\end{equation}
Differentiating Eqs. (\ref{dsta}) and (\ref{dste}) again, we obtain
\begin{equation}
\ddot{S}_{TA}=\frac{9\pi}{2G}(1+\omega)^2\left[1+16\pi G \Omega _m \left(\frac{1+3\omega}{1-3\omega}\right)-\frac{m_{g}^{2}}{H_c}\left(3\beta R_A-\frac{\gamma}{H_c}\right)\right]
\end{equation}
and
\begin{eqnarray}
\ddot{S}_{TE}&=&\frac{\pi}{G}\Big[\left\lbrace (1-3\omega)H^2 R_{E}^{2}-6HR_E+2\right\rbrace -\frac{3m_{g}^{2}}{2H_c}\left(\frac{1+\omega}{1-3\omega}\right)\Big\lbrace \left(3\beta R_A-\frac{2\gamma}{H_c}+\frac{3\delta}{H_{c}^{2}}\frac{1}{R_A}\right)
\nonumber
\\
&\times &\left( (1-3\omega)H^2 R_{E}^{2}-2HR_E-2\right) +9(1+\omega)\left(\beta -\frac{\delta}{H_{c}^{2}}\frac{1}{R_{A}^{2}}\right)R_E (HR_E+1)\Big\rbrace 
\nonumber
\\
&-&96\pi G \Omega _m\left(\frac{1+\omega}{1-3\omega}\right)(HR_E-1)\Big].
\end{eqnarray}

As the expressions for the first and second order time variation of the total entropy are very complicated, so it is not possible to predict the validity of GSLT and thermodynamical equilibrium (given by inequalities (\ref{tecases})) analytically. Hence we have examined their validity from the graphical representation in Figs. 1-4 for various choices of the parameters involved.

\begin{figure}
\begin{minipage}{0.4\textwidth}
\includegraphics[width=1.0\linewidth]{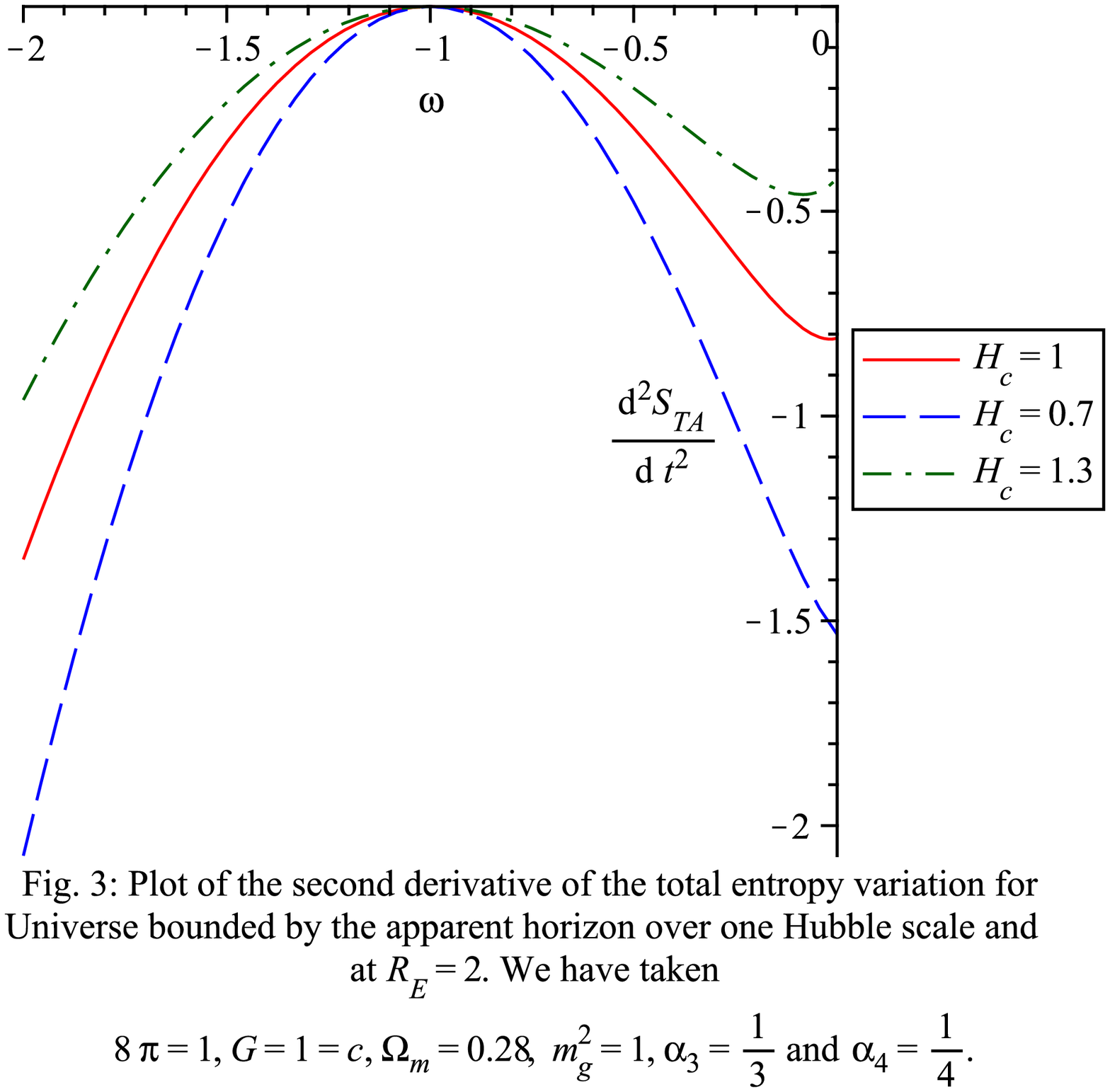}
\end{minipage}
\begin{minipage}{0.4\textwidth}
\includegraphics[width=1.0\linewidth]{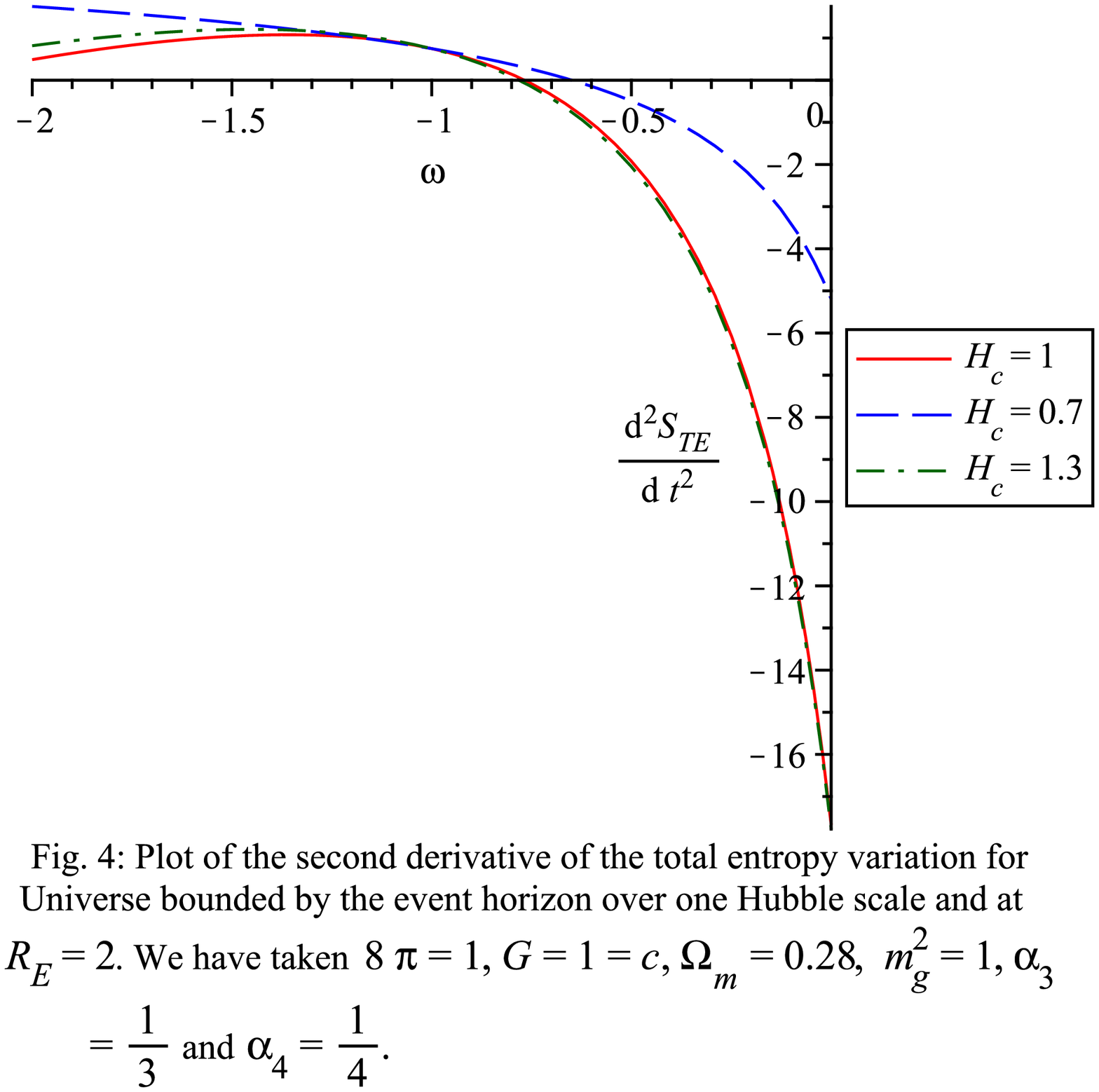}
\end{minipage}
\end{figure}

\section{Brief Discussion and Final Comments}

This paper deals with the study of Universal thermodynamics, particularly the generalized second law of thermodynamics (GSLT) and the thermodynamical equilibrium (TE) for Universe bounded by an apparent/event horizon in massive gravity under the assumption that the first law (i.e., Clausius relation) holds. Cosmological solutions of massive gravity theory as a viable alternative to describe the present day cosmological observations have been widely investigated. But unfortunately, it was found \cite{De Felice1} that all homogeneous and isotropic solutions in dGRT theory are unstable. Moreover, there does not exist any nontrivial FLRW solution with Minkowskian reference metric. So in the present work, we have chosen the reference metric to be de Sitter for which the theory does admit flat FLRW solutions (see Eq. (\ref{ft1})) but it is unstable in nature. Thus the aim of this work was to analyze this cosmological system in the thermodynamical perspective. We have established modified forms for entropy (Eqs. (\ref{sa1}) and (\ref{se1})) associated with the apparent/event horizon using methods proposed by Hayward and Cai. Using these equations and the Gibb's equation, we have been able to evaluate the total entropy variation and its derivative for Universe bounded by the apparent/event horizon and filled with perfect fluid having constant equation of state parameter $\omega$. In Figs. 1 and 2, variation of the total entropy has been plotted against $\omega$ and ($H_c=1, 0.7, 1.3$) for Universe bounded by the apparent and the event horizon respectively over one Hubble scale and at $R_E=2$. The other parameters have been chosen as $\Omega _m=0.28$, $m_{g}^{2}=1$, $\alpha _3=\frac{1}{3}$ and $\alpha _4=\frac{1}{4}$. We see that GSLT holds for the apparent horizon when the fluid is not beyond $\Lambda$CDM while for the event horizon, GSLT holds when the fluid is almost exotic in nature (i.e., $\omega <-\frac{1}{3}$). On the other hand, Figs. 3 and 4 depict the TE for Universe bounded by the apparent and the event horizon respectively for similar values of the parameters as in Figs. 1 and 2. It is evident from Fig. 3 that TE holds unconditionally for the apparent horizon except at $\omega =-1$, where we have the limiting situation. Fig. 4 shows that validity of TE for the event horizon is restricted only for normal fluid. Therefore, based on the above analysis, we can conclude that in the present unstable cosmological scenario (in massive gravity theory), the apparent horizon is more favourable as compared to the event horizon from thermodynamical perspective in massive gravity, which is in contrast to our earlier observations \cite{Saha1}.

\begin{acknowledgments}

The authors are thankful to IUCAA, Pune, India for their warm hospitality and research facilities as the work has been done there during a visit. The author SM is thankful to UGC, Govt. of India for providing NET-JRF. Author SS is thankful to UGC-BSR Programme of Jadavpur University for awarding research fellowship. Author SC acknowledges the UGC-DRS Programme in the Department of Mathematics, Jadavpur University. 

\end{acknowledgments}

\section*{Appendix A}

Using Eq. (\ref{pnew}) in Eq. (\ref{sa}), one obtains
\begin{eqnarray}
S_A &=& \frac{A_A}{4G}-\frac{\pi}{G} \int m_{g}^2 \frac{\dot{H}}{H^2}\frac{H}{H_c}\left[-3\beta +2\gamma \frac{H}{H_c}-3\delta \frac{H^2}{H_{c}^{2}}\right]HR_{A}^{4}dt \nonumber \\
&=& \frac{A_A}{4G}-\frac{\pi m_{g}^2}{G H_c} \int \left[-3\beta +2\gamma \frac{H}{H_c}-3\delta \frac{H^2}{H_{c}^{2}}\right]\frac{dH}{H^4} \nonumber \\
&=& \frac{A_A}{4G}-\frac{\pi m_{g}^2}{G H_c} \left[\beta R_{A}^{3} -\frac{\gamma}{H_c}R_{A}^{2}+\frac{3\delta}{H_{c}^{2}}R_{A}\right], \nonumber
\end{eqnarray}
which is Eq. (\ref{samg}).\\\\
Using Eq. (\ref{pnew}) in Eq. (\ref{se}), the entropy of the event horizon takes the form
\begin{eqnarray}
S_E &=& \frac{A_E}{4G}-\frac{\pi}{2G} \int \left(\frac{R_{A}^{2}R_E}{1-\epsilon}\right)\left(\frac{HR_E+1}{HR_E-1}\right) \left[m_{g}^2 \frac{\dot{H}}{H^2}\frac{H}{H_c} \left(-3\beta +2\gamma \frac{H}{H_c}-3\delta \frac{H^2}{H_{c}^{2}} \right)\right] dR_E \nonumber \\
&=& \frac{A_E}{4G}-\frac{\pi m_{g}^2}{2GH_c} \int R_E \left(\frac{HR_E+1}{1-\epsilon} \right)\left[-3\beta +2\gamma \frac{H}{H_c}-3\delta \frac{H^2}{H_{c}^{2}}\right] \frac{dH}{H^3} \nonumber \\
&=& \frac{A_E}{4G}-\frac{\pi m_{g}^2}{2GH_c} \int \left[ \frac{2R_E(HR_E+1)}{1-\dot{R_A}} \left\lbrace \frac{-3\beta}{H^3}+\frac{2\gamma}{H^2 H_c}-\frac{3\delta}{H H_{c}^{2}}\right\rbrace \right] dH \nonumber ,
\end{eqnarray}
which is Eq. (\ref{semg}).

\section*{Appendix B}

From Eq. (\ref{fhe}),
\begin{eqnarray}
T_{f} dS_{fh} &=& dE_f+pdV_h \nonumber \\
or,~~ T_{f} \frac{dS_{fh}}{dt} &=& \dot{\rho}_m V_h+(\rho _m+p_m)\frac{dV_h}{dt} \nonumber \\
or,~~ T_{f} \dot{S}_{fh} &=& -3H(\rho _m +p_m) \cdot \frac{4}{3}\pi R_{h}^{3}+4\pi (\rho _m+p_m)R_{h}^{2}\dot{R}_h \nonumber \\
or,~~ \dot{S}_{fh} &=& \frac{4\pi R_{h}^{2}}{T_f} (\rho _m+p_m)(\dot{R}_h-HR_h), \nonumber
\end{eqnarray}
which is Eq. (\ref{dfh}).

\end{document}